\documentstyle[preprint,prl,aps]{revtex}

\begin{document}

\draft

\title{GUT effects in the soft supersymmetry breaking terms}

\author{Nir Polonsky and Alex Pomarol}
\address{Department of Physics, University of Pennsylvania,
Philadelphia, Pennsylvania, 19104, USA}
\date{May 1994, UPR-0616T}

\def\calo{{\cal O}}
\def\call{{\cal L}}
\def\calm{{\cal M}}
\def\ie{{\it i.e.}}
\def\eg{{\it e.g.}}
\def\be{\begin{equation}}
\def\ee{\end{equation}}
\def\bea{\begin{eqnarray}}
\def\eea{\end{eqnarray}}
\def\mpla{M_{P}}
\def\mgut{M_{G}}
\def\tr{{\rm tr}}
\def\hone{{\cal H}_{1}}
\def\htwo{{\cal H}_{2}}
\def\mssm{\rm SU(3)_c\times SU(2)_L\times U(1)_Y}

\maketitle

\begin{abstract}
In minimal supergravity theories
the soft supersymmetry breaking (SSB) parameters
are universal (flavor blind)
near the Planck scale.
Nevertheless,  one often assumes universality at
the grand-unification  scale $\mgut \approx
10^{16}$ GeV instead, and corrections
to the SSB parameters arising from their  evolution
between the Planck and GUT scales are neglected.
We study these corrections and show
that large splittings between the scalar mass parameters
can be induced at $M_{G}$.
These effects are model dependent
and lead to significant uncertainties in
the low-energy predictions of supersymmetric
models, in their correlations and in the allowed
parameter space.
\end{abstract}

The minimal supersymmetric extension of the standard model (MSSM)
is a well motivated candidate to describe the physics
beyond the standard model \cite{review}.
The unknown origin of supersymmetry
breaking is parametrized by soft supersymmetry
breaking (SSB) terms in the lagrangian
(that do not reintroduce quadratic divergencies),
$i.e.$,
\be
-\call_{soft}=m^2_i|\Phi_i|^2+B_{ij}\Phi_i\Phi_j
+A_{ijk}\Phi_i\Phi_j\Phi_k
+\frac{1}{2}M_{\alpha}\lambda_{\alpha}^2+h.c.\, ,
\label{softg}
\ee
where $\Phi_i$ ($\lambda_{\alpha}$)
are the scalar (gaugino) fields.
Eq.~(\ref{softg}) introduces
a large number of new arbitrary parameters
and is impractical for phenomenological studies.
A better  situation
appears if the supersymmetry is a local symmetry, \ie, supergravity.
It is then assumed that
supersymmetry is broken in a hidden sector
which  couples only  gravitationally to the observable sector.
The effective lagrangian for the observable sector below
$\mpla\equiv M_{Planck}/\sqrt{8\pi} \approx 2.4 \times 10^{18}$ GeV
consists of a global supersymmetric theory
with SSB terms as in eq.~(\ref{softg}).
In the minimal supergravity model,
which we assume here, the K\"ahler potential is
flat
and one finds that the SSB parameters have universal values at $M_{P}$
\cite{review}, \ie,
\be
m^2_i\equiv m^2_0,\ \ B_{ij}\equiv B_0\mu_{ij},\ \
A_{ijk}\equiv A_0Y_{ijk},\ \ M_{\alpha}\equiv M_{1/2}\, ,
\label{softuni}
\ee
where $\mu_{ij}$ and $Y_{ijk}$ are respectively
the bilinear and trilinear couplings in the superpotential.
The deviations
from the universal boundary condition (\ref{softuni})
at lower scales are
calculated using renormalization group (RG) methods, and
given only four
soft parameters  one can predict the superpartner
mass spectrum.

If the MSSM is embedded
in a grand-unified theory (GUT) at the
scale $M_{G} \approx 10^{16}$ GeV
suggested by coupling constant unification
\cite{ccunification},
then the evolution of the parameters between $M_{P}$
and $M_{G}$ depends on the GUT and
is strongly model dependent.
Nevertheless, it is often assumed that applying (\ref{softuni})
at $M_{G}$ rather than at $M_{P}$  is a good approximation,
because $M_{G}$ is close to $M_{P}$.
One then uses the generic MSSM
RG equations (RGEs) between $M_{G}$ and the
weak scale \cite{examples,penn594}.

In this letter we will
examine the corrections
to the SSB parameters arising
from their evolution
between the Planck and the GUT scales.
We will show that these
corrections induce large deviations in the SSB parameters
from their universal values.
The corrections  are typically
proportional to
${\alpha\over \pi}m^{2} \ln{M_{P}/M_{G}}$
where $\alpha$ and $m^{2}$ are
a generic coupling and soft mass parameter, respectively.
Although these corrections are not enhanced by large
logarithms, they can be significant due to:
\begin{enumerate}
\item The number of particles above  $\mgut$, $N$,
is large as a result of the large symmetry group, and
one roughly
has ${\alpha\over \pi} \rightarrow  {N\alpha\over \pi}$.
(See also Ref. \cite{earlywork}.)
\item Large Yukawa couplings that are typically present in GUTs
and that
grow with the energy.
In addition
to the large top Yukawa coupling,
one has  to  introduce
extra large couplings to avoid a too large proton decay rate.
\end{enumerate}
Corrections from the gauge sector
[$\propto \frac{\alpha_{G}}{\pi}M_{1/2}^{2} \ln{M_{P}/M_{G}}$]
can also be important for large   $M_{1/2}$ or
if $\alpha_{G}$ grows with the energy
(as in non-minimal GUTs).
The above corrections depend on the details of the GUT model
and represent
uncertainties in the low-energy predictions.
Gravitational and other effects could also
affect the boundary condition (\ref{softuni})
and would only add to the uncertainty.

For definiteness and simplicity we consider
the minimal SU(5) model. We will comment on extended
models below.
The Higgs sector of the model consists of three supermultiplets,
$\Sigma({\bf 24})$ in the adjoint representation [which is
responsible for the breaking of SU(5) down to $\mssm$], and
$\hone ({\bf\bar 5})$ and $\htwo ({\bf 5})$, each containing
a SU(2) doublet
$H_i$ and a color triplet $H_{C_i}$.
The matter superfields are in the ${\bf\bar 5}+{\bf 10}$ representations,
$\phi({\bf\bar 5})$ and $\psi({\bf 10})$. The superpotential is given
by\footnote{We define $\Sigma=\sqrt{2}T_{a}w_{a}$ where
$T_a$ are the SU(5) generators with $\tr\{T_aT_b\}=\delta_{ab}/2$, and
we only consider Yukawa couplings for the third generation.}
\bea
W&=&\mu_\Sigma\tr\Sigma^{2}+\frac{1}{6}\lambda^{'}\tr\Sigma^{3}
+\mu_H\hone\htwo +\lambda \hone\Sigma \htwo\nonumber \\
&+&\frac{1}{4}h_t\epsilon_{ijklm}\psi^{ij}\psi^{kl}\htwo^m
+\sqrt{2}h_b\psi^{ij}\phi_{i}{\cal H}_{1j}\, .
\label{super}
\eea
In the supersymmetric limit
$\Sigma$ develops a
vacuum expectation value
$\langle\Sigma\rangle =\nu_{\Sigma}\, {\rm diag}
(2,2,2,-3,-3)$ and
the  gauge bosons  $X$ and $Y$ get a mass
$M_V=5g_{G}\nu_{\Sigma}$.
In order for
the Higgs SU(2) doublets to have masses  of $\calo(m_Z)$ instead
of $\calo(\mgut)$,
the fine-tuning
$\mu_H-3\lambda \nu_{\Sigma} \lesssim \calo(m_Z)$
is required and one obtains
$M_{H_{C}}=\frac{\lambda}{g_G} M_V$.
Dimension-five operators
induced by the color triplet give large contributions
$\propto 1/M^2_{H_C}$
to the proton decay rate \cite{proton}.
To suppress such operators, the mass of the color triplets has to
be large, $M_{H_{C}}\gtrsim M_V$,
implying
$\lambda\gtrsim g_G\approx 0.7$.
Thus, one-loop corrections proportional to $\lambda$
produce important effects.
Below $\mpla$ the effective lagrangian also contains the SSB terms
\bea
-\call_{soft}&=&m^2_{\hone}|\hone|^2+m^2_{\htwo}|\htwo|^2
+m^2_{\Sigma}\tr\{\Sigma^{\dag}\Sigma\}
+m^2_5|\phi|^2+m^2_{10}\tr\{\psi^{\dag}\psi\} \nonumber \\
&+&[B_\Sigma \mu_\Sigma\tr\Sigma^2
+\frac{1}{6}A_{\lambda^{'}}
\lambda^{'}\tr\Sigma^3
+B_H\mu_H\hone\htwo
+A_{\lambda}\lambda\hone\Sigma\htwo\nonumber \\
&+&\frac{1}{4}A_th_t\epsilon_{ijklm}\psi^{ij}\psi^{kl}\htwo^m
+\sqrt{2}A_bh_b\psi^{ij}\phi_{i}{\cal H}_{1j}+
\frac{1}{2}M_5\lambda_{\alpha}
\lambda_{\alpha}+h.c.]\, .
\label{soft}
\eea
 From
$\mpla$ to $\mgut$ the SSB terms evolve according to the
RGEs of the SU(5) model with eq.~(\ref{softuni}) as a boundary condition.
Thus,
we  expect
a breakdown of universality at $\mgut$
for SSB parameters
of fields that are in different SU(5) representations.
The SU(5) RGEs for the
SSB parameters and Yukawa couplings
are given by
\bea
\frac{dm^2_{10}}{dt}&=&\frac{1}{8\pi^2}[3h^2_t(m^2_{\htwo}+2m^2_{10}
+A^2_t)+2h^2_b(m^2_{\hone}+m^2_{10}+m^2_{5}+A^2_b)
-\frac{72}{5}g_G^2M_{5}^2]\, ,\nonumber \\
\frac{dm^2_{5}}{dt}&=&\frac{1}{8\pi^2}[4h^2_b(m^2_{\hone}+m^2_{10}
+m^2_{5}+A^2_b)-\frac{48}{5}g_G^2M_{5}^2]\, ,\nonumber \\
\frac{dm^2_{\hone}}{dt}&=&\frac{1}{8\pi^2}[4h^2_b(m^2_{\hone}+m^2_{10}
+m^2_{5}+A^2_b)
+\frac{24}{5}\lambda^2(m^2_{\hone}+m^2_{\htwo}+m^2_{\Sigma}+A^2_{\lambda})
-\frac{48}{5}g^2_GM_{5}^2]\, ,\nonumber \\
\frac{dm^2_{\htwo}}{dt}&=&\frac{1}{8\pi^2}[3h^2_t(m^2_{\htwo}+2m^2_{10}
+A^2_t)
+\frac{24}{5}\lambda^2(m^2_{\hone}+m^2_{\htwo}+m^2_{\Sigma}+A^2_{\lambda})
-\frac{48}{5}g^2_GM_{5}^2]\, ,\nonumber \\
\frac{dm^2_{\Sigma}}{dt}&=&\frac{1}{8\pi^2}[\frac{21}{20}
\lambda^{\prime 2}
(3m^2_{\Sigma}+A^2_{\lambda '})+\lambda^2(m^2_{\hone}+m^2_{\htwo}
+m^2_{\Sigma}+A^2_\lambda)
-20g^2_GM_{5}^2]\, ,\nonumber \\
\frac{d\lambda '}{dt}&=&\frac{\lambda '}{16\pi ^2}[\frac{63}{20}
\lambda^{\prime 2}+3\lambda^2-30g^2_G]\, ,\ \ \
\frac{d\lambda }{dt}=\frac{\lambda }{16\pi ^2}[\frac{21}{20}
\lambda^{\prime 2}+3h^2_t+4h^2_b+\frac{53}{5}\lambda ^2
-\frac{98}{5}g^2_G]\, ,\nonumber \\
\frac{dh_t}{dt}&=&\frac{h_t}{16\pi ^2}[9h^2_t+4h^2_b
+\frac{24}{5}\lambda^2-\frac{96}{5}g^2_G]\, ,\ \ \
\frac{dh_b}{dt}=\frac{h_b}{16\pi ^2}[10h^2_b+3h^2_t+\frac{24}{5}\lambda^2
-\frac{84}{5}g^2_G]\, ,
\label{rgesu5}
\eea
where $t=\ln Q$.
The RGE for the gauge coupling is
$d\alpha_G/dt=-3\alpha_G^2/2\pi$, and
similarly $dM_{5}/dt=-3\alpha_G M_{5}/2\pi$.
The RGEs for the trilinear SSB parameter $A_i$ can be obtained from the
RGEs of the corresponding Yukawa coupling $Y_i$ by
\be
\frac{dY_i}{dt}=\frac{Y_i}{16\pi^2}\left[a_{ij}Y^2_j-bg^2_G\right]\rightarrow
\frac{dA_i}{dt}=\frac{1}{8\pi^2}\left[a_{ij}Y^2_jA_j-bg^2_GM_5\right]\, .
\ee
We can omit the RGEs for $\mu_\Sigma$, $\mu_H$, $B_\Sigma$ and $B_H$,
which are arbitrary parameters that decouple from the rest
of the RGEs.

The evolution of the SSB parameters from
$M_{P}$
to $\mgut$ is dictated by a
competition between the  positive Yukawa terms
($i.e.$, scalar contributions) and the
negative gauge terms ($i.e.$, gaugino contributions) in the RGEs.
We can distinguish two scenarios:
(A) For moderate values of $M_{1/2}\equiv M_5(M_{P})$
the contribution from the gauge sector
is small.
In this case,
the RGEs of $m^2_{\hone}$ and $m^2_{\htwo}$ have a large contribution
proportional to $\lambda^2$
and  both masses are diminished
as the energy scale decreases. For $h_t\gg h_b$,
$m^2_{\htwo}$ decreases faster than  $m^2_{\hone}$ but also  $m^2_{10}$
(for the third family) is diminished in that case.
(B) For large values of $M_{1/2}$ the RGEs are dominated by the negative
gaugino contribution so that all the SSB parameters increase as the
energy scale decreases.
The scalar masses  are enhanced by an additive factor
\be
\Delta m^2_{i}=-\frac{c_i}{3}\left[1-\frac{1}
{\left(1+\frac{3\alpha_G}{2\pi}\ln\frac{M_{G}}{M_{P}}\right)^2}
\right]M^2_{1/2}\, ,
\label{shift}
\ee
where $c_i=\frac{72}{10}(\frac{24}{5})$ for $i$ in the ${\bf 10}({\bf 5})$
representation.
One has $\Delta m^2_{i}\approx 0.5(0.3)M^2_{1/2}$.

Examples of scenarios A and B are given in Figs.~1a and 1b, respectively.
We see that
the violation of the universality of the SSB parameters at $\mgut$
can be substantial.
In particular, the soft masses of the Higgs fields are typically split
from the matter field masses.
For $M_{H_C}=1.4M_V$ ($i.e.$, $\lambda \approx 1$ at $\mgut$)
the splitting can be as large as
$100\%$.

In order to analyze the implications of these soft mass splittings
in the
 supersymmetric spectrum and phenomenology,
we have
to run  the SSB parameters from $\mgut$ down to $m_Z$
\cite{ack}.
Below $\mgut$ the effective theory corresponds to the MSSM:
\be
W=\mu H_1H_2+h_tQH_2U+h_bQH_1D+h_{\tau}LH_1E\, ,
\label{effsuper}
\ee
where $Q$($L$) and $U$, $D$($E$) are respectively
the quark (lepton) SU(2) doublet and
singlet superfields.
The tree-level
matching conditions of the SSB parameters between the SU(5) model
and the MSSM are
$m^2_{{\cal H}_{i}}(\mgut)=m^2_{H_{i}}(\mgut)$,
$m^2_{10}(\mgut)=m^2_{Q,\,U,\,E}(\mgut)$ and
$m^2_{5}(\mgut)=m^2_{D,\,L}(\mgut)$.
One-loop matching conditions will be considered below.
In this letter we present only a qualitative analysis of the
GUT effects in the low energy quantities.
A comprehensive numerical study,
together with the details of the numerical procedures,
will be given elsewhere.

In scenario A
the parameters $m^2_{{\cal H}_i}$ have substantial shifts
and  the parameters $m^2_{H_{i}}(m_{Z})$
are modified. The latter enter the
minimization conditions of the weak-scale  Higgs potential.
Therefore,
corrections to $m^2_{H_{i}}$
modify the region of the MSSM parameter space that
is consistent with electroweak symmetry breaking (EWSB).
The  $\mu$ parameter
is also affected by the GUT corrections to $m^2_{H_i}$
since it is
extracted from
the minimization
conditions of the weak-scale Higgs potential\cite{review,examples,penn594}.
This leads to corrections to
observables
that depend on $\mu$, such as
the Higgsino mass,
the Higgsino-gaugino
mixing and the  left-right scalar quark mixing.
Also, consistency with
constraints from color and charge breaking \cite{penn594}
can be affected.
A priori, one would expect
that the  Higgs boson masses are also
affected. Notice, however,
that the latter
depend only on the sum $m^2_{H_{i}}+|\mu|^2$, which is not modified
significantly
(the shift in  $m^2_{H_{i}}$ is approximately compensated by the shift in
$|\mu|^2$.)
The masses of the  lightest chargino and neutralinos
 are typically proportional to $M_{1/2}$,
and are only slightly affected.
There are also large deviations
for the scalar quark masses of  the third generation.
As mentioned above, due to the evolution from $M_{P}$ down to $\mgut$,
$m_{10}^2$ can  be shifted
for large $h_t$ ($e.g.$, $\tan\beta=\langle H_2 \rangle/ \langle H_1 \rangle
\approx 1$)
from the universal value $m^2_0$
(see Fig.~1a). In addition, the evolution
of $m^2_{Q}$ and $m^2_{U}$
from $\mgut$ down to $m_Z$ depends on the value of $m^2_{H_{2}}$ that
is sensitive to the GUT physics.

In scenario B, where
the gaugino contribution is dominant,
all the scalar masses are shifted
[eq.~(\ref{shift})] from
their universal value at $M_{P}$.
The scalar quark squared masses, however, receive large corrections
in the running from $\mgut$ down to $m_Z$ ($\approx 6M^2_{5}$) so that
the amount (\ref{shift})
represents only a few percent of their values.
This is not the case for the scalar leptons, where such corrections  are
much smaller [$\approx
0.5(0.1)M^2_5$ for $\tilde e_L(\tilde e_R)$] and
the increment (\ref{shift}) can even double their masses.

In Figs.~2  and 3 we present examples of the GUT effects
in the low-energy predictions.
In Fig.~2 we take a large $\tan\beta$
($\tan\beta = 42$) and show the allowed values
(\ie, consistent with EWSB and experimental bounds)
of $\mu$ $vs.$ the gluino mass, $M_{\tilde g}$,
 when the evolution from $\mpla$ to $\mgut$ is
considered (triangles) or neglected (filled circles).
The correlation between $\mu$ and $M_{\tilde g}$, which exists
when neglecting the $M_{P}$ to $M_{G}$ evolution \cite{examples,penn594},
is ``smeared'' when that evolution is included, and
$\mu$ is larger in the latter case. We also find that
$m_{H_{i}}^{2}(m_{Z})$ are often both negative
when the GUT effects are considered
[$m^{2}_{{\cal H}_{i}}(M_{G})$ are diminished together
because $h_{t} \approx h_{b}$],
which is inconsistent
with EWSB. The allowed parameter space is then significantly reduced.
In Fig.~3 we show the
light $t$-scalar mass $m_{\tilde{t}_1}$ $vs.$
$M_{\tilde g}$  for $\tan\beta \approx 1$
($h_{t} \approx 1$).
$\mu$ is now large ($\sim 1$ TeV) and is less sensitive to the
GUT effects. Corrections to
$m_{\tilde{t}_1}$
are mainly via the diminished
$h_{t}^{2} m_{H_{2}}^{2}$ term
in the respective RGEs below $M_{G}$.
$\tilde{t}_{1}$ is therefore heavier
and some points which correspond to a tachionic
$t$-scalar and are excluded  when  the $M_{P}$ to $M_{G}$ evolution is
neglected, can be allowed.
Note also that the correlation between $m_{\tilde{t}_1}$ and $M_{\tilde g}$ is
weakened by the GUT corrections. We find that
correlations between predictions are genericly modified due
to the model-dependent ``smearing'' from the $M_{P}$ to $M_{G}$
evolution. The correlation
between
$m_{\tilde{t}_{1}}$  and $m_{\tilde{t}_{2}}$ is, however, strengthened
because of the heavier $\tilde{t}_{1}$.
Figs.~\ref{fig:fig2} and \ref{fig:fig3} correspond to scenario B (A)
for very large (small to moderate)
values of $M_{\tilde g}$.

Even if the universal boundary condition (\ref{softuni}) for the
SSB parameters is taken at
$\mgut$, there is some arbitrariness in the value of $\mgut$ due to
mass-splittings between  the particles at the GUT scale, \ie, threshold
effects.
We will distinguish
three categories of corrections
(details will be  given elsewhere).
First, we consider logarithmic threshold corrections
arising from the mass splitting between  different heavy superfields.
Threshold corrections to Yukawa and gauge couplings
are discussed, for example, in Ref. \cite{us}.
The largest contributions to the SSB parameters
arise from the SU(2) triplet and singlet components of the $\Sigma$ superfield,
$\Sigma_{3}$ and $\Sigma_{1}$,
 and are given by (we identify $\mgut = \max{\{M_V,\, M_{H_C}\}}$)
\begin{mathletters}
\label{logthresholdb}
\begin{equation}
m^2_{H_{i}}(\mgut) = m^2_{{\cal H}_i}(\mgut)+
\frac{\lambda^2}{4\pi^2}(m^2_{\hone}
+m^2_{\htwo}+m^2_{\Sigma}+A^2_{\lambda})
\left[\frac{3}{4}\ln\frac{M_{\Sigma_{3}}}{\mgut}
+\frac{3}{20}\ln\frac{M_{\Sigma_{1}}}{\mgut}\right]\, ,
\end{equation}
\begin{equation}
\Delta{A_{t,b}( \mgut)} = \frac{\lambda^2}{4\pi^2}A_{\lambda}
\left[\frac{3}{4}\ln\frac{M_{\Sigma_{3}}}{\mgut}+\frac{3}{20}
\ln\frac{M_{\Sigma_{1}}}{\mgut}\right]\, .
\end{equation}
\end{mathletters}
Since the masses $M_{\Sigma_{3}}$ and
$M_{\Sigma_{1}} \equiv 0.2M_{\Sigma_{3}}$
can be much smaller than $\mgut$,
these corrections can be substantial.
For $M_{\Sigma_3}\approx 10^{-2}\mgut$ and $\lambda\approx 1$,
we have $m^2_{H_{i}}(\mgut)\approx 0.6m^2_0$.
The second type of threshold effects are logarithmic corrections
due to the boson-fermion mass splitting within a supermultiplet.
Such corrections are suppressed by powers of $m_{soft}/\mgut$
in  the Yukawa and gauge coupling boundary conditions,
but there is no such suppression for the SSB mass terms,
\eg, corrections to $m^2_{H_i}$
are $\sim \mgut^2\ln\left[(\mgut^2+m^2_{soft})/\mgut^2\right]\sim
m^2_{soft}$.
Keeping only the terms of $\calo(\lambda^2/4\pi^2)$, we have
\be
m^2_{H_{i}}(\mgut)=m^2_{{\cal H}_i}(\mgut)+
\frac{\lambda^2}{4\pi^2}\left[\frac{18}{20}
(m^2_{\Sigma}
+A_{\lambda}B_{\Sigma})+\frac{6}{8}
(m^2_{\hone}+m^2_{\htwo}+2A_{\lambda}B_H)\right]\, ,
\label{logthresholdc}
\ee
where the first term comes from corrections of the
$\Sigma_{3}$ and $\Sigma_{1}$ particles, while the second term
from corrections of the Higgs color triplets.
This represents a $\calo (10\%)$ correction.
Lastly, there are scheme-dependent finite one-loop corrections.
In the dimensional-reduction scheme they are given by
\be
m^2_{H_{i}}(\mgut)=m^2_{{\cal H}_i}(\mgut)-
\frac{\lambda^2}{4\pi^2}\frac{24}{20}(m^2_{\hone}+
m^2_{\htwo}+m^2_{\Sigma}
+A^2_{\lambda})\, .
\label{logthresholdd}
\ee
Notice that the corrections (\ref{logthresholdc})
tend to cancel the corrections
(\ref{logthresholdd}) for equal SSB parameters.
 From (\ref{logthresholdb}) -- (\ref{logthresholdd})
one expects an additional  $\calo (40\%)$ common correction to
$m_{H_{i}}^{2}(M_{G})$ that would induce $\calo (5-20\%)$ uncertainties
in low-energy predictions.

In extended supersymmetric GUTs one expects the corrections to be larger.
If large representations are introduced,
 the positive  scalar contribution to the RGEs is larger
and therefore the SSB parameters decrease faster with the scale.
However, one has to be aware of a possible
breakdown of  perturbation theory.
An  interesting scenario occurs
in models in which $\hone$ and $\htwo$ couple with different
strength to the other Higgs supermultiplets.
For example, in the  missing partner SU(5) model
 $W=\lambda_1\hone\Sigma ({\bf 75})\Phi ({\bf 50})+
\lambda_2\htwo\Sigma ({\bf 75})\Phi ({\bf\bar{50}})+\dots$,
and if $\lambda_2>\lambda_1$,  the evolution  from $\mpla$ to $\mgut$
splits the two Higgs scalar masses.
That  splitting  can  now affect
the low-energy Higgs boson masses
and  reduce
the degree of fine-tuning  that is typically required to achieve EWSB
in scenarios with large $\tan\beta$
(in which the Higgs masses are not split
by Yukawa interactions).
If we enlarge the symmetry group, the negative term in the RGEs
coming from the gaugino contribution is enhanced and can partially
cancel the scalar contribution.
In models where the rank of the group is larger than the
rank of the SM group, \eg, SO(10), we have an additional contribution to the
scalar masses that arises from the D-terms \cite{dterms}.

To summarize, we have shown that large deviations from universality at
$M_{G}$ can be generated when considering ($i$) the model-dependent evolution
from $M_{P}$ to $M_{G}$ and ($ii$) threshold corrections
at $M_{G}$ (including those from
scalar-fermion splittings).
We have also shown that the above leads to a modification of the allowed
parameter space, smears predicted correlations and affects certain
low-energy predictions such as the $\mu$ parameter and
 the $t$-scalar mass.
These corrections have to be considered as
 uncertainties  when analyzing possible future
evidence for supersymmetry. On the other hand, such corrections
could provide a probe of the high scale.

\acknowledgments
 It is a pleasure to thank P. Langacker for discussions and comments
on the manuscript, and M. Cvetic for discussions.
This work was supported by the US Department of Energy
Grant No. DE-AC02-76-ERO-3071 (NP) and by the Texas Commission
Grant No. RGFY93-292B (AP).

\begin{figure}
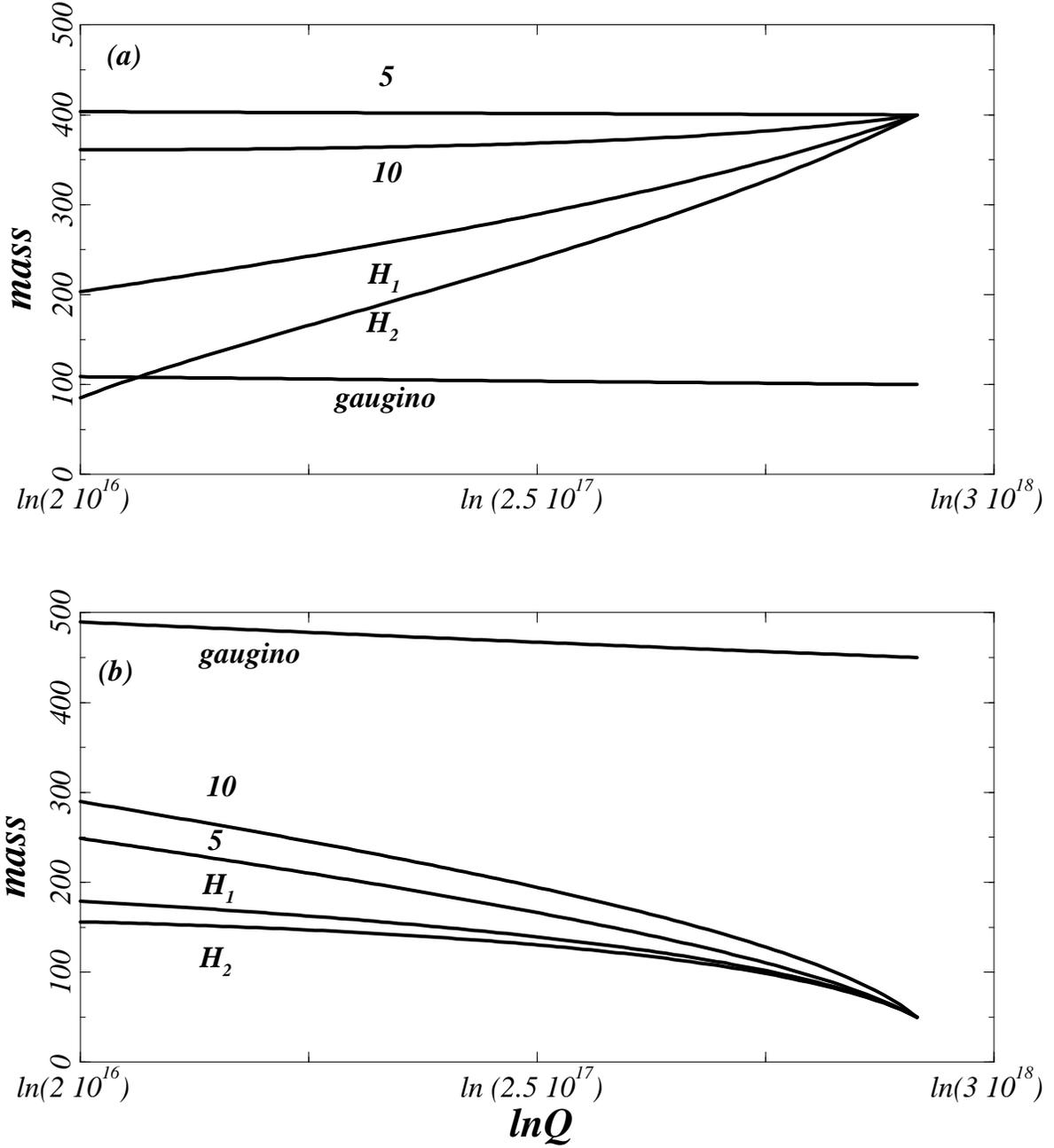

\caption{The evolution of the soft mass parameters of the third family,
 $\phi({\bf \bar{5}})$ and $\psi({\bf 10})$, the
 Higgs, ${\cal H}_{i}$, and the gaugino
between the Planck and grand-unification
scales for:
($a$) scenario A, with
$m_{0}=A_{0}= 400$ GeV and $M_{1/2} = 100$ GeV;
and
($b$) scenario B, with
$m_{0}=A_{0}= 50$ GeV and $M_{1/2} = 450$ GeV.
In both cases
$m_{t} = 160$ GeV;  $\tan\beta = 1.25$; and
the boundary conditions $\lambda =1$
($i.e.$, $M_{H_{C}} = 1.4M_{V}$)
and $\lambda^{'} = 0.1$
at $M_{G}$ are assumed. All masses are in GeV.
}
\label{fig:fig1}
\end{figure}
\begin{figure}
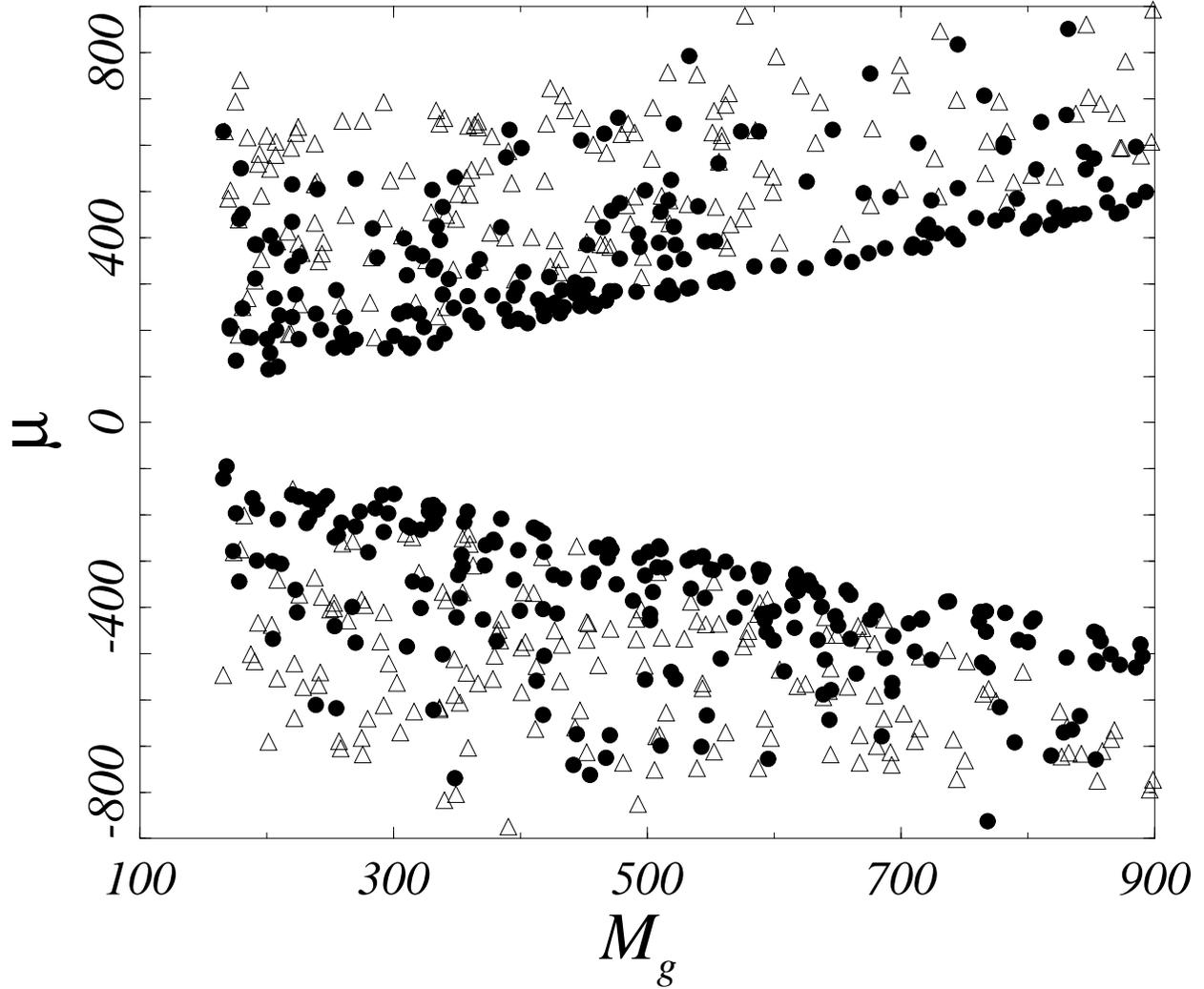

\caption{Scatter plot
of the $\mu$ parameter
$vs.$ the
gluino mass
within the allowed parameter space
for $m_{t} = 180$ GeV and $\tan\beta = 42$.
Triangles (filled circles) correspond to
universality [eq.~(2)]
at the Planck (grand-unification)
scale. $\lambda$ and $\lambda^{'}$ are as in Fig.~1. All masses
are in GeV.
}
\label{fig:fig2}
\end{figure}
\begin{figure}
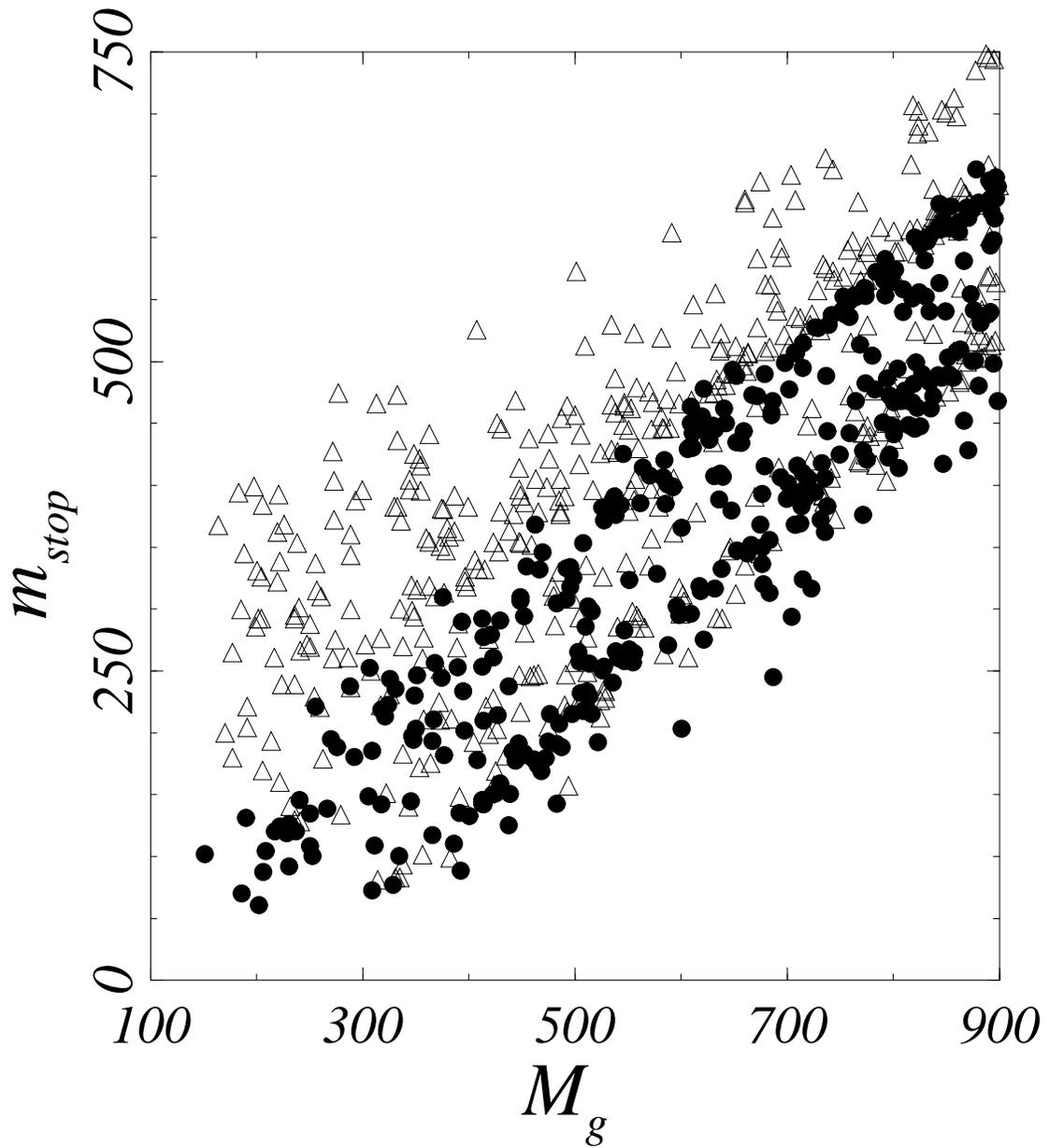

\caption{Same as in Fig.~2 except
 the light $t$-scalar mass
$vs.$ the
gluino  mass, and
$m_{t} = 160$ GeV and $\tan\beta = 1.25$.}
\label{fig:fig3}
\end{figure}

\end{document}